\documentstyle[12pt]{article}
\begin{document}

\topmargin 0pt
\oddsidemargin 7mm
\headheight 0pt
\topskip 0mm
\addtolength{\baselineskip}{0.6\baselineskip}

\thispagestyle{empty}

\rightline{SOGANG-HEP-200/95}
\rightline{hep-th/9506166}
\rightline{May 1995}

\begin{center}
\vspace{36pt}
{ \large \bf On the anomaly of nonlocal symmetry in the
chiral QED}
\end{center}

\vspace{36pt}

\begin{center}
Hyeonjoon Shin, Young-Jai Park, and Yongduk Kim
\end{center}
\begin{center}
{\it Department of Physics$^*$ and Basic Science Research Institute,
Sogang University, C.P.O. Box 1142, Seoul 100-611, Korea}
\end{center}
\vspace{1cm}
\begin{center}
Won T. Kim
\end{center}
\begin{center}
{\it  Department of Science Education and Basic Science Research Institute, \\
Ehwa Women's University, Seoul 120-750, Korea}
\end{center}
\vfill

\begin{center}
{\bf ABSTRACT}
\end{center}

We show that the anomaly of nonlocal symmetry can be canceled by the
well-known Wess-Zumino acton which cancels the gauge anomay in
the two-dimensional chiral electrodynamics.
\vspace{1cm}
\\
------------------------------------------- \\
$^*$e-mail address: hshin, yjpark, wtkim@physics.sogang.ac.kr\\
\newpage

Quantum gauge symmetries are very important to build modern
quatum field theories. Recently, Lavelle and McMullan [1] have
shown that there exists a distinct nonlocal symmetry which is
different from the usual gauge symmetry in the
quantum electrodynamics (QED). The original gauge symmetry is
implemented by the Becci-Rouet-Stora-Tyutin (BRST) symmetry [2]
in the gauge fixed action.
The nonlocal symmetry, which the gauge fixing term is invariant,
is dual to the BRST symmetry.
The generalized Lorentz covariant symmetry has discussed in Ref.[3].
On the other hand, the classical gauge symmetry or BRST symmetry
may be spoiled by the anomaly
after quantization in the chiral QED, which has a chiral gauge coupling [4].
Therefore, it is natural to study whether the nonlocal symmetry
may be anomalous in chiral QED or not.

In this Brief Report,
we show that the anomaly of the nonlocal symmetry exists, and
this can be canceled by the well-known Wess-Zumino action [5] in the exactly
solvable two-dimensional chiral QED.

The chiral QED is defined by the gauge invariant Lagrangian as follows
\begin{equation}
{\cal L}_{0}= - \frac{1}{4} F_{\mu\nu} F^{\mu\nu}
          +i \bar{\psi} \gamma^\mu (\partial_\mu -ie A_\mu \frac{1-\gamma_5}
          {2} ) \psi,
\end{equation}
and we choose the Lorentz invariant gauge
fixing term and the corresponding ghost action,
\begin{equation}
{\cal L}_{\rm GF}+{\cal L}_{\rm Ghost}
= -\frac{1}{2\alpha} (\partial_\mu A^\mu)^2  -i \partial_\mu \bar{c}
   \partial^\mu c.
\end{equation}
Then, the total classical action ${\cal L}_{\rm C}={\cal L}_{0}
+{\cal L}_{\rm GF}+ {\cal L}_{\rm Ghost}$ is BRST invariant under the following
transformations as
\begin{eqnarray}
&\delta_B \psi = ie c \psi ,~~~ \delta_B \bar{\psi}=ie\bar{\psi} c &
\nonumber \\
& \delta_B A_\mu = -\frac{1}{e}\partial_\mu c,~~~ \delta_B c = 0,~~~
\delta_B \bar{c} = \frac{i}{e \alpha} \partial_\mu A^\mu. &
\end{eqnarray}
It is also invariant under the nonlocal transformation [1] as follows
\begin{eqnarray}
& & \delta^\perp \psi = e \left( \frac{\partial_0}{\nabla^2}c \right) \psi,
 ~~~\delta^\perp \bar{\psi} = e \bar{\psi} \frac{\partial_0}{\nabla^2}\bar{c} ,
 \nonumber \\
& & \delta^\perp A_0 = i \bar{c}, ~~~
    \delta^\perp A_1 = i \frac{\partial_1 \partial_0}{\nabla^2}
    \bar{c}, \label{rule} \\
& & \delta^\perp \bar{c} = 0, ~~~
    \delta^\perp c = - A_0 + \frac{\partial_1 \partial_0}{\nabla^2} A_1 +
                     \frac{e}{\nabla^2} j_0 ~, \nonumber
\end{eqnarray}
where $j^0 = e \bar{\psi} \gamma^0 \frac{1-\gamma_5}{2} \psi$.

Let us now consider the quantum effective action to investigate
the anomaly of the nonlocal symmetry at the quantum level.
The effective action, which incorporates one fermionic loop [4], is exactly
calculated as
\begin{equation}
{\cal L}_{\rm eff} = \frac{1}{2} a e^2 A_\mu A^\mu -\frac{e^2}{2} A_\mu
        \frac{(g^{\mu\nu} +\epsilon^{\mu\nu})\partial_\nu \partial_\rho
       (g^{\rho\sigma}-
       \epsilon^{\rho\sigma})}{\Box}   A_\sigma ~,
\end{equation}
where the constant $a$ is a regularization ambiguity.
Then the BRST anomaly of the total quantum effective action,
${\cal L}_{\rm Q}={\cal L}_{\rm eff}+{\cal L}_{\rm GF}+{\cal L}_{\rm Ghost}$
is given by
\begin{equation}
\delta_B {\cal L}_{\rm Q} = e\left[  \epsilon^{\mu\nu} \partial_\mu A_\nu +
(a-1)  \partial_\mu A^\mu \right] c.
\end{equation}

On the other hand,
the nonlocal transformations are defined by exactly the same forms in
Eq.~(\ref{rule}) except for the ghost field $c$ by the following rule
\begin{equation}
\delta^\perp c = - A_0 + \frac{\partial_1 \partial_0}{\nabla^2} A_1 +
                     \frac{e}{\nabla^2} J_0 ~,
\end{equation}
where $ J_0 = ae A_0 - e (\partial_0+\partial_1) \frac{1}{\Box}
          (g^{\mu\nu}-\epsilon^{\mu \nu}) \partial_\mu A_\nu $.
Then, the anomaly of the nonlocal transformation is given by
\begin{equation}
\delta^\perp {\cal L}_Q = i e^2 \bar{c} \frac{\partial_0}{\nabla^2}
    \left[ \epsilon^{\mu \nu} \partial_\mu A_\nu
 +(a-1) \partial_\mu A^\mu     \right].
\end{equation}
It is interesting to note that the nonlocal symmetry can be broken
after quantization similar to the BRST case in the two-dimensional chiral QED.
To recover this symmetry, it is sufficient to add the following Wess-Zumino
action ${\cal L}_{\rm WZ}$ to the effective action ${\cal L}_{\rm Q}$ [5],
\begin{equation}
{\cal L}_{\rm WZ} =\frac{1}{2} (a-1) \partial_\mu \theta \partial^\mu \theta
+ e \left[ (a-1) g^{\mu \nu} +\epsilon^{\mu \nu} \right] \partial_\mu \theta
A_\nu  .
\end{equation}
Then, after the integration of the $\theta$ field, the resulting
Lagrangian is given by
\begin{equation}
{\cal L} = - \frac{1}{4} F_{\mu \nu} {\Box}^{-1} (\Box +m^2) F^{\mu \nu}
-\frac{1}{2\alpha} (\partial_\mu A^\mu)^2  - i
                     \partial^\mu \bar{c} \partial_\mu c ~,
\end{equation}
where $m^2 =\frac{ae^2}{a-1}$ for $a >1$.
This Lagrangian is manifestly invariant under the nonlocal transformation
as well as the BRST transformation.

In conclusion, we have shown that the nonlocal symmetry may be anomalous after
quantization in the chiral QED. However, it is sufficient
to add the well-known Wess-Zumino action to recover the nonlocal symmetry.
It would be interesting to study the nonlocal anomaly related to the
chiral anomaly in higher dimensions and its geometrical meaning.

\section*{ACKNOWLEDGMENTS}
The present study was supported by Ministry of Education, Project No.
BSRI-95-2414 (1995).


\begin{thebibliography}{99}
\bibitem{1} M. Lavelle and David McMullan, Phys. Rev. Lett.
           {\bf 71}, 3758 (1993).
\bibitem{2} T. Kugo and I. Ojima, Prog. Theor. Phys. Suppl. {\bf 66}, 1(1979).
\bibitem{3} Z. Tang and D. Finkelstein, Phys. Rev. Lett. {\bf 73},
           3055 (1994).
\bibitem{4} R. Jackiw and Rajaraman, Phys. Rev. Lett. {\bf 54},
           1219 (1985).
\bibitem{5} K. Harada and I. Tsutsui, Phys. Lett. {\bf B 183}, 311 (1987).
\end{thebibliography}
\end{document}